\documentclass[fleqn,usenatbib]{mnras}
\usepackage{newtxtext,newtxmath}
\usepackage[T1]{fontenc}
\usepackage{ae,aecompl}
\usepackage[usenames,dvipsnames]{xcolor}
\usepackage{graphicx}
\usepackage{amsmath}
\usepackage{amssymb}
\usepackage{siunitx}
\usepackage{microtype}
\usepackage{indentfirst}
\usepackage[T1]{fontenc}
\usepackage{graphicx}
\usepackage{epstopdf}
\usepackage{booktabs}
\usepackage{bibentry}
\usepackage{amsmath}
\usepackage{color}
\usepackage{ulem}
\usepackage{hyperref}
\usepackage{setspace}

\hypersetup{colorlinks,linkcolor=blue,citecolor=blue}

\graphicspath{{./}{figures/}}  

\DeclareSIUnit\MHz{\mega\hertz}
\DeclareSIUnit\kHz{\kilo\hertz}
\DeclareSIUnit\jansky{Jy}
\DeclareSIUnit\mJy{\milli\jansky}
\DeclareSIUnit\MJy{\mega\jansky}
\DeclareSIUnit\mK{\milli\kelvin}
\DeclareSIUnit\parsec{pc}
\DeclareSIUnit\arcsec{arcsec}
\DeclareSIUnit\Mpc{\mega\parsec}

\newcommand{\degree}{$^{\circ}$}

\newcommand{\Ha}{H$\alpha$~}

\title[Contamination of Galactic free--free emission]{%
  Contribution of Galactic free--free emission to the foreground for EoR signal in SKA experiments
}

\author[Lian~et~al.]{%
Xiaoli Lian,$^{1}$ \thanks{E-mail:
	\href{mailto:lianxiaoli87@sjtu.edu.cn}{lianxiaoli87@sjtu.edu.cn} (LX)} Haiguang Xu,$^{1,2,3}$ Zhenghao Zhu,$^{1}$ Dan Hu$^{1}$
\\
$^{1}${School of Physics and Astronomy,
  Shanghai Jiao Tong University,
  800 Dongchuan Road, Shanghai 200240, China} \\
$^{2}${Tsung-Dao Lee Institute, Shanghai Jiao Tong University, 800 Dongchuan Road, Shanghai 200240, China}\\
$^{3}${IFSA Collaborative Innovation Center, Shanghai Jiao Tong University, 800 Dongchuan Road, Shanghai 200240, China}\\
}

\date{%
}

\pubyear{2020}

\begin{document}
\label{firstpage}
\pagerange{\pageref{firstpage}--\pageref{lastpage}}
\maketitle

%
%
\begin{abstract}
	
The overwhelming foreground contamination hinders the accurate detection of the 21-cm signal of neutral hydrogen during the Epoch of Reionization (EoR). Among various foreground components, the Galactic free--free emission is less studied, so that its impact on the EoR observations remains unclear. In this work, we employ the observed \Ha intensity map with the correction of dust absorption and scattering, the {\sc Simfast21} software, and the latest SKA1-Low layout configuration to simulate the SKA ``observed'' images of Galactic free--free emission and the EoR signal. By calculating the one-dimensional power spectra from the simulated image cubes, we find that the Galactic free--free emission is about \numrange{e3.5}{e2.0}, \numrange{e3.0}{e1.3}, and \numrange{e2.5}{e1.0} times more luminous than the EoR signal on scales of $\SI{0.1}{\per\Mpc} < k < \SI{2}{\per\Mpc}$ in the \numrange{116}{124}, \numrange{146}{154}, and \SIrange{186}{194}{\MHz} frequency bands. We further analyse the two-dimensional power spectra inside the properly defined EoR window and find that the leaked Galactic free--free emission can still cause non-negligible contamination, as the ratios of its power (amplitude squared) to the EoR signal power can reach about \num{200}\%, \num{60}\%, and \num{15}\% on scales of \SI{1.2}{\per\Mpc} in three frequency bands, respectively. Therefore, we conclude that the Galactic free--free emission, as a severe contaminating foreground component, needs to be carefully treated in the forthcoming deep EoR observations. 
 
\end{abstract}



 \section{Introduction}
\label{chap:introduction}

\noindent A phase transition occurred when the first stars and galaxies began to form after the Dark Ages ($z \sim \numrange{30}{200}$) and the Cosmic Dawn ($z \sim \numrange{15}{30}$). During this phase transition, the UV photons emitted from the first ionizing sources (e.g., first stars, galaxies, and quasars) reionized most of the hydrogen in the ambient intergalactic medium (IGM), so that the corresponding period is named the Epoch of Reionization (EoR; $z \sim \numrange{6}{15}$) (e.g., \citealt{Furlanetto06,Morales10,Loeb12,Koopmans15,Furlanetto16}). The ionized regions are expected to appear as bubbles, which gradually grew larger and finally merged. Statistical information about the expansion of ionized bubbles and these astrophysical sources are encoded in the power spectrum of EoR signal, which is a statistical measurement of the fluctuations in $k$-space (e.g., \citealt{Sims16}). Among various EoR probes, the redshifted 21-cm hyperfine emission line is expected to be the most promising one \citep{Furlanetto16}, which provides a wealth of information about both the first ionizing sources and the ionization states of IGM \citep{Cooray04,Bharadwaj04}. Theoretical models have predicted that the brightness temperature of the EoR signal is about the order of \SI{10}{\mK} (e.g., \citealt{Mesinger11}). However, the overwhelming foreground contamination is the primary obstacle to accurately detect the extremely weak EoR signal.

During the past decade several low-frequency radio interferometers have been designed and constructed to probe the EoR signal, including the LOw Frequency ARray (LOFAR; \citealt{van13})\footnote{\url{http://www.lofar.org/}}, the Giant Metrewave Radio Telescope (GMRT; \citealt{Paciga13})\footnote{\url{http://www.gmrt.ncra.tifr.res.in}}, the Murchison Widefield Array (MWA; \citealt{Tingay13})\footnote{\url{http://www.mwatelescope.org/}}, the 21 Centimetre Array (21CMA; \citealt{Wang10}), and the Donald C. Backer Precision Array for Probing the Epoch of Reionization (PAPER; \citealt{Parsons10})\footnote{\url{http://eor.berkeley.edu/}}. The next-generation instruments such as the Hydrogen Epoch of Reionization Array (HERA; \citealt{DeBoer17})\footnote{\url{http://reionization.org/}} and the Square Kilometer Array (SKA; \citealt{Mellema13,Koopmans15})\footnote{\url{https://www.skatelescope.org/}} have also been in construction to achieve larger collecting area and higher sensitivity. However, there are still a variety of challenges to be addressed in these instruments, such as the ionospheric distortions, the frequency artifacts, and the \numrange{3}{5} orders of magnitude overwhelming foreground contamination \citep{Shaver99,Morales10}.

The majority of foreground contamination is contributed by the Galactic synchrotron emission (\num{\sim 70}\%) and extragalactic point sources (\num{\sim 27}\%; \citealt{Shaver99,DiMatteo04,Murray17,Spinelli18}). Among various foreground components, the Galactic free--free emission ($\lesssim \num{1}\%$) is less studied, so that its impact on the EoR signal is still poorly understood. At intermediate and high Galactic latitudes, the Galactic synchrotron component dominates the emission at frequencies lower than \SI{10}{\GHz}, the dust thermal emission becomes overwhelming at frequencies higher than \SI{100}{\GHz}, whereas the Galactic free--free emission becomes important in \SIrange{10}{100}{\GHz} frequency band and may reach the comparable levels to the cosmic microwave background (CMB) fluctuations and other foregrounds (e.g., anomalous microwave emission: AME, synchrotron emission, etc.) depending on sky position \citep{Planck-XXIII15,Planck-XXV15,Planck-IX16,Planck-X16,PlanckXIII16}. However, the \Ha emission line (the \numrange{3}{2} transition of the hydrogen atom at $\lambda = \num{656.28}$ nm) provides a novel way to trace the Galactic free--free emission away from the Galactic plane, i.e., intermediate and high Galactic latitudes, since they share the same emission measure $\rm{EM}$ $\equiv \int n_{\rm e}^2 dl$ ($n_{\rm e}$ is the electron density; e.g., \citealt{Dickinson03}). For example, the brightness temperature of Galactic free--free emission has been related to the \Ha intensity by \citet{Valls98}, and a Galactic free--free emission map at \SI{30}{\GHz} covering \num{95}\% of the sky (except the area $|b| < \SI{5}{\degree}$, $l = \SI{160}{\degree}$--\SI{0}{\degree}--\SI{260}{\degree}) has been proposed by \citet{Dickinson03}. Note that the above results may have been biased since the observed \Ha intensities used in these works are often misunderstood, due to dust absorption and scattering, especially in the Galactic plane (e.g., \citealt{Dennison98,Dickinson03}). In this work, we derive a more reliable all-sky Galactic free--free emission map by applying the correction of dust absorption and scattering to the observed \Ha intensity map. Meanwhile, we employ the {\sc Simfast21} code to simulate the brightness temperatures of the EoR signal. To incorporate the instrumental effects, the latest SKA1-Low layout configuration is considered. By analysing the power spectra and the EoR window (implying foreground removal and avoidance methods; e.g., \cite{Chapman16,Sims16,Li19}), we quantitatively evaluate the contamination caused by Galactic free--free emission on the EoR detection. 

This paper is structured as follows. We present the calculations of the foreground components including the Galactic free--free emission, Galactic synchrotron emission, and extragalactic point sources in Section \ref{chap:foregrounds}. In Section \ref{chap:21cm}, we simulate the EoR signal by using the new version of {\sc Simfast21} software. In Section \ref{chap:sim-obs}, we employ the latest SKA1-Low layout configuration to incorporate the instrumental effects to the simulated the SKA ``observed'' images of foreground components and the EoR signal. We briefly introduce the power spectra and EoR window in Section \ref{chap:ps-eor}, and then quantitatively evaluate the contamination caused by Galactic free--free emission on the EoR signal in Section \ref{chap:results}. In Section \ref{chap:dis}, we present the impacts of sky positions and frequency artifacts on the EoR detection and show the effect of Galactic free--free emission on the component separation. Finally, we summarize our work in Section \ref{chap:summary}. Throughout this work a $\Lambda$CDM cosmology is assumed with parameters of $\Omega _{m} = \Omega_{dm} + \Omega_{b} = \num{0.3089}$, $\Omega_{b} = \num{0.0486}$, $\Omega_{\Lambda} = \num{0.6911}$, Hubble constant $H_0 = \SI{67.74}{\km\per\second\per\Mpc}$, power spectrum index $n_{s} = \num{0.9667}$, and the normalisation $\sigma_8 = \num{0.8159}$ \citep{PlanckXIII16}. 

\section{Simulation of Foreground Components}
\label{chap:foregrounds}

We simulate the low-frequency radio sky by considering the Galactic free--free emission, Galactic synchrotron emission, and extragalactic point sources, with an emphasis on the Galactic free--free emission. We choose the sky map centered at ($\rm R.A.$, $\rm Dec.$) = (\SI{0}{\degree}, \SI{-30}{\degree}) with a sky coverage of \SI[product-units=repeat]{10 x 10}{\degree}, which locates at a high galactic latitude $({l, ~b})$ = (\SI{16}{\degree}, \SI{-78}{\degree}) and is expected to be an appropriate choice for this study (region A in Figure \ref{fig:maps}). Moreover, this region passes through the zenith of the SKA1-Low telescope and is an ideal choice to simulate the SKA observation (more details see Section \ref{chap:sim-obs}). Each simulated sky map is divided into \num{1800} $\times$ \num{1800} pixels with a pixel size of \SI{20}{\arcsecond} in Cartesian projection. We choose three representative frequency bands of \numrange{116}{124}, \numrange{146}{154}, and \SIrange{186}{194}{\MHz}, with a frequency bandwidth of \SI{8}{\MHz} to limit the cosmological evolution in the calculation of power spectrum of the EoR signal (e.g., \citealt{Thyagarajan13,Li19}). For each frequency band, the \SI{8}{\MHz} bandwidth is divided into \num{51} channels with a frequency resolution of \SI{160}{\kHz} to carry out the calculations. 

\subsection{Galactic Free--free Emission}
\label{chap:free-free}

We derive the Galactic free--free emission from the observed \Ha intensity map. \citet{Finkbeiner03} released an all-sky \Ha intensity map\footnote{\url{https://faun.rc.fas.harvard.edu/dfink/skymaps/}} at a resolution of \SI{6}{\arcminute} by combining three \Ha surveys, i.e., Virginia Tech Spectral line Survey (VTSS\footnote{\url{http://www.phys.vt.edu/~halpha}}; \citealt{Dennison98}), Southern H-Alpha Sky Survey Atlas (SHASSA\footnote{\url{http://amundsen.swarthmore.edu/SHASSA}}; \citealt{Gaustad01}), and the northern sky survey of Wisconsin H-Alpha Mapper (WHAM\footnote{\url{http://www.astro.wisc.edu/wham}}; \citealt{Haffner03}). In \num{2009}, The WHAM was moved to Cerro Tololo in Chile to observe the southern sky to complete a full-sky \Ha survey with approximately \SI{1}{\degree} resolution, known as Wisconsin H-Alpha Mapper Sky Survey (WHAM-SS) \citep{Haffner10}. Given the higher spatial resolution, we employ the F03 \Ha intensity map, incorporating the $E(B-V)$ color map and the $100$ ${\rm \mu m}$ infrared intensity map \citep{Schlegel98} that have been used to correct the dust absorption and scattering, to deduce the Galactic free--free emission map. 	

\subsubsection{Dust-correction of \Ha Intensity}
\label{halpha}

\citet{Bennett03} proposed a method to correct the absorption of the F03 \Ha intensity map $I_{\rm H\alpha}^{\rm obs}(\textbf{r})$

\begin{equation}
I_{{\rm H\alpha}}^{\rm a-corr}(\textbf{r}) = I_{\rm H\alpha}^{\rm obs}(\textbf{r})~~\tau_{\rm H\alpha}(\textbf{r})~/~(1 - e^{-\tau_{\rm H\alpha}(\textbf{r})}),
\label{eq:dust-extinction}
\end{equation}

\noindent where $I_{\rm H\alpha}^{\rm obs}(\textbf{r})$ ($\textbf{r}$ is the two-dimensional position) is the observed \Ha intensity in units of Rayleigh ($\rm R$)\footnote{$1$ Rayleigh $\rm{(R)} \equiv 10^{6}/4\pi~$$\rm{photons~s^{-1}~cm^{-2}~sr^{-1}}$ $\equiv 2.41 \times 10^{-7} \rm{erg~cm^{-2}~s^{-1}~sr^{-1}}$, $I_{{\rm H\alpha}}^{\rm a-corr}(\textbf{r})$ is the absorption-corrected \Ha intensity in units of $\rm {R}$}, and $\tau_{\rm H\alpha}(\textbf{r})$ is the optical depth at \Ha wavelength that can be derived based on the $E_{(B-V)}(\textbf{r})$ color map provided by \citet{Schlegel98} (e.g., $\tau_{\rm H\alpha}(\textbf{r}) = 2.2~E_{(B-V)}(\textbf{r})$; \citealt{Bennett13}). 

Furthermore, we apply an approximate scattering correction method \citep{Witt10,Brandt12} to correct the scattering of \Ha intensity based on the dust \num{100} ${\rm \mu m}$ infrared emission 

\begin{equation}
I_{\rm H\alpha}^{\rm s-corr}(\textbf{r}) = I_{\rm H\alpha}^{\rm obs}(\textbf{r})~-~0.11~I_{\rm 100~\mu m}(\textbf{r}),
\label{eq:dust-scattering}
\end{equation}

\noindent where $I_{\rm H\alpha}^{\rm s-corr}(\textbf{r})$ is the scattering-corrected \Ha intensity in units of $\rm {R}$, $I_{\rm 100~\mu m}(\textbf{r})$ is the \num{100} ${\rm \mu m}$ infrared intensity of \citet{Schlegel98} in units of \si{\MJy\per\steradian}, and the coefficient \num{0.11} is the mean values given by \citet{Witt10} ($0.129 \pm 0.015~\rm R / MJy~sr^{-1}$) and \citet{Brandt12} ($0.090 \pm 0.017~\rm R / MJy~sr^{-1}$). Although the correlation was measured in regions where $\tau_{\rm H\alpha}(\textbf{r}) < 1$, we apply this relation over the entire sky by following \citet{Bennett13}, for the regions at intermediate and high Galactic latitudes, on which we will focus in this work, meet the requirements of optically thin. In the analyzed \SI[product-units=repeat]{10 x 10}{\degree} region, the \Ha intensity is about \numrange{0.5}{1.0} $\rm{R}$, and the \num{100} ${\rm \mu m}$ infrared intensity shows a mean value of \num{0.8} $\rm MJy~sr^{-1}$, so that the scattering-corrected effect is about \num{9}\%--\num{18}\%. In the top panels of Figure \ref{fig:maps}, we present the observed F03 \Ha intensity map, the $E_{(B-V)}(\textbf{r})$ map, and the \num{100} ${\rm \mu m}$ infrared intensity $I_{\rm 100~\mu m}(\textbf{r})$ map. It is clearly shown that the emissions of $I_{\rm H\alpha}^{\rm obs}(\textbf{r})$, $E_{(B-V)}(\textbf{r})$, and $I_{\rm 100~\mu m}(\textbf{r})$ are concentrated on the Galactic plane, which show typical values of $\num{0.8}~\rm {R}$, $\num{0.1}~\rm {mag}$, and \SI{1.0}{\MJy\per\steradian} at high latitudes, respectively.

\subsubsection{Distribution of Galactic Free-free Emission}

\label{chapter:ha-to-ff}

The received \Ha intensity depends on whether the emitting medium is optically thin (case A) or optically thick (case B), and it is found that case B is satisfied in the study of Galactic \Ha emission \citep{Osterbrock89}. For case B, \citet{Valls98} proposed an analytical expression to describe the relationship between the observed \Ha intensity $I_{\rm H\alpha}^{\rm obs}(\textbf{r})$ and the emission measure ${\rm EM}(\textbf{r})$

\begin{equation}
\label{equ:EM}
{\rm EM}(\textbf{r}) = 2.561~~ T_{4}^{1.017}(\textbf{r}) ~~10^{0.029/T_{4}(\textbf{r})}~~ I_{\rm H\alpha}^{\rm obs}(\textbf{r}) ~~ [\rm cm^{-6}~pc],
\end{equation}

\noindent where ${\rm EM}(\textbf{r})$ is in units of $\rm cm^{-6}~pc$, and $T_{4}(\textbf{r})$ = $T_{\rm e}(\textbf{r}) / 10^4$ is the electron temperature in units of \SI{e4}{K}. Using Equation \ref{equ:EM} the optical depth of Galactic free--free emission $\tau_{\rm c}(\textbf{r})$ can be derived by 

\begin{equation}
\label{equ:tau-Draine}
\tau_{\rm c}(\textbf{r}) = 0.05468~~g(\textbf{r})~~T_{\rm e}(\textbf{r})^{-3/2}~~ \nu_{9}^{-2}~~ {\rm EM}(\textbf{r}),
\end{equation}

\noindent where $T_{\rm e}(\textbf{r})$ is in units of \si{\K}, $\nu_{9}$ = $\nu / \SI{e9}{\Hz}$ is the observed frequency in units of \si{\GHz}, and $g(\textbf{r})$ is the gaunt factor that can be derived by  

\begin{equation}
\label{equ:g-Draine}
g(\textbf{r}) = {\rm log} \{{\rm exp[5.960 - \sqrt{3}}/\pi {\rm log}(\nu_{9}(\textbf{r}) T_{4}(\textbf{r})^{-3/2} )] + \rm e\},
\end{equation}	

\noindent where $\rm e = \num{2.71828}....$ is the natural constant \citep{Draine11}. The above three Equations are valid in the frequency range of \SI{100}{\MHz}--\SI{100}{\GHz} and are often quoted to deduce the Galactic free--free emission, which can be expressed in brightness temperature as


\begin{equation}
\label{equ:tb}
T_{\rm b}^{\rm {Gff}}(\textbf{r}) = T_{\rm e}(\textbf{r})~~[1 - {\rm exp}(- \tau_{\rm c}(\textbf{r}))],
\end{equation}

\noindent where $T_{\rm b}^{\rm Gff}(\textbf{r})$ is the brightness temperature of Galactic free--free emission, and $\tau_{\rm c}(\textbf{r})$ is given by Equation \ref{equ:tau-Draine}. 

\begin{figure*}
	\centering
	\includegraphics[width=0.32\textwidth]{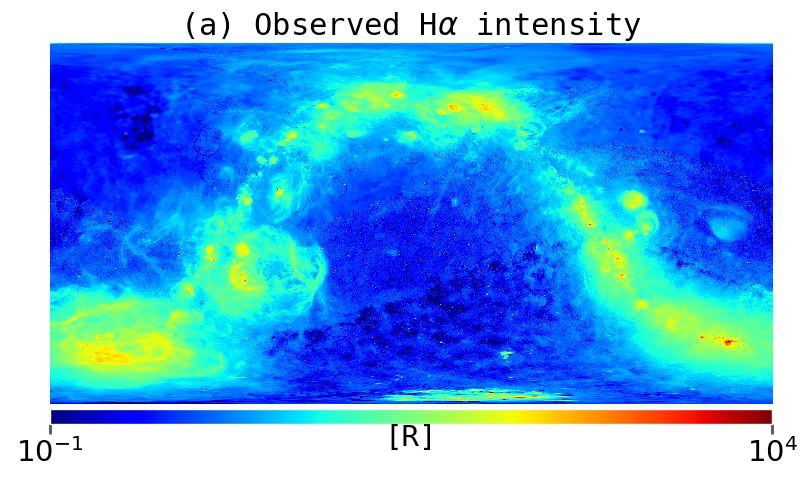}
	\includegraphics[width=0.32\textwidth]{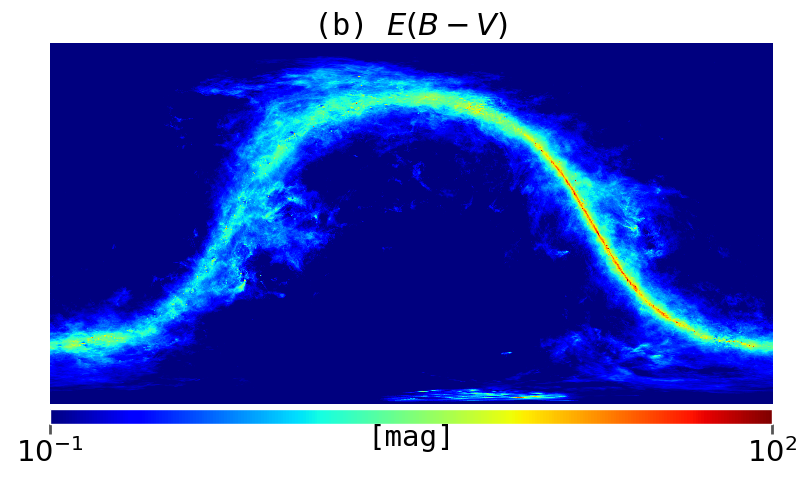}
	\includegraphics[width=0.32\textwidth]{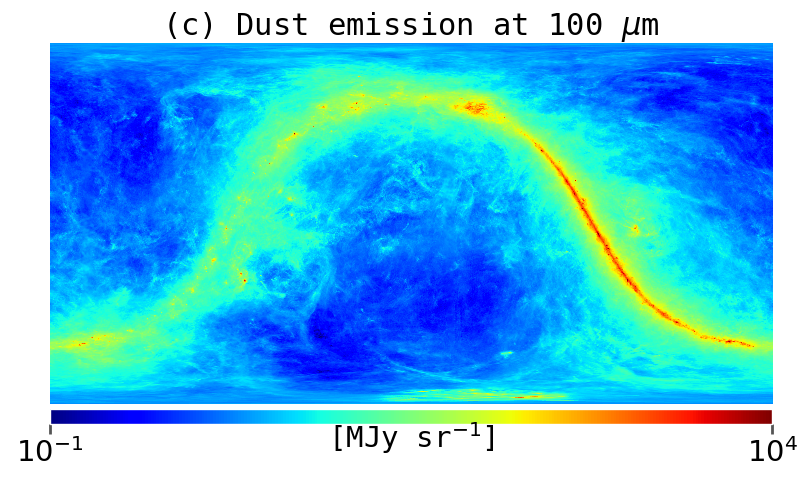}
	\centering
	\includegraphics[width=0.32\textwidth]{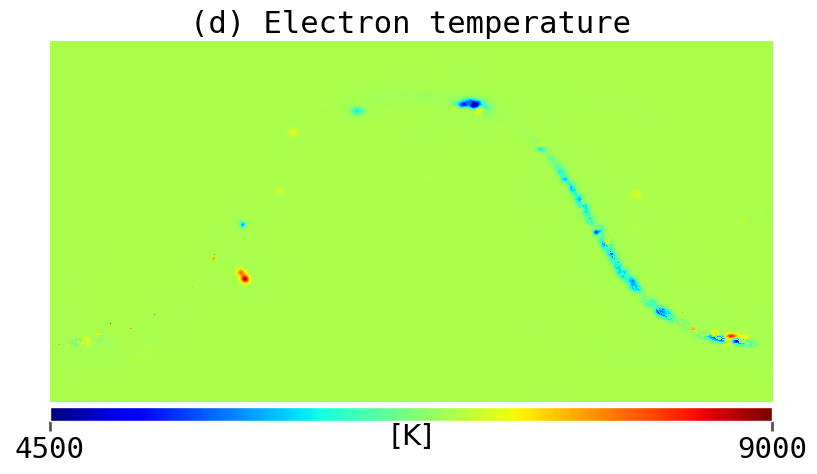}
	\includegraphics[width=0.32\textwidth]{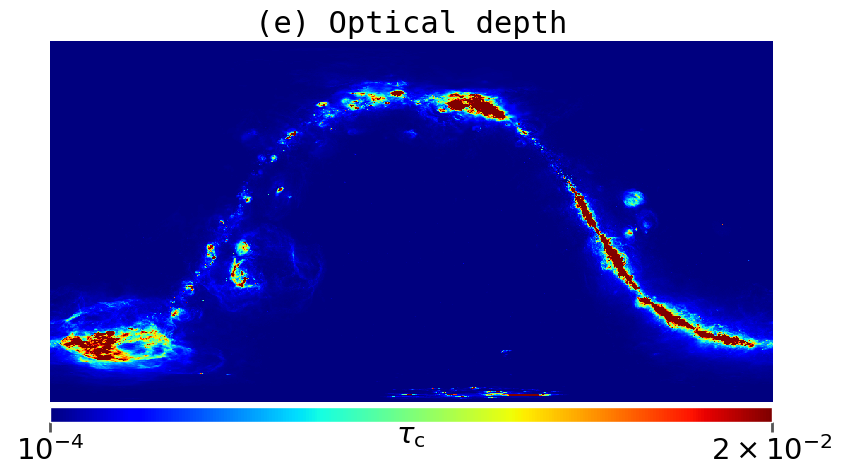}
	\includegraphics[width=0.32\textwidth]{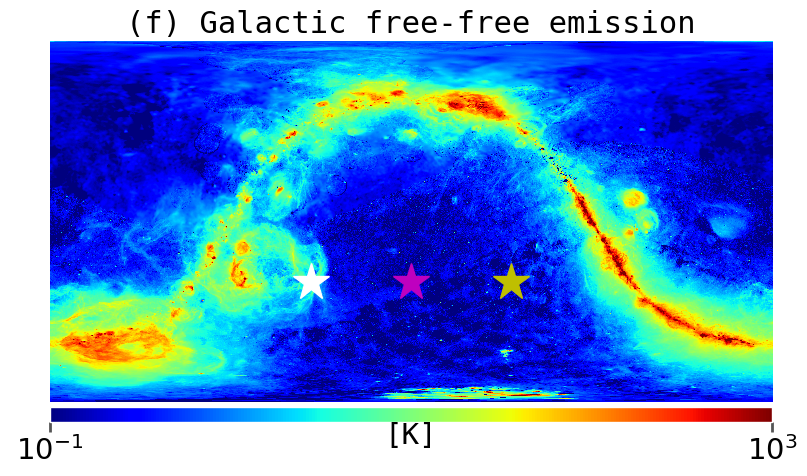}
	\caption{The all-sky maps of (a) the observed F03 \Ha intensity $I_{\rm H\alpha}^{\rm obs}(\textbf{r})$, (b) $E(B-V)$ \citep{Schlegel98}, (c) \num{100} ${\rm \mu m}$ \citep{Schlegel98}, (d) electron temperature $T_{\rm e}(\textbf{r})$ \citep{Planck-X16}, (e) optical depth $\tau_{\rm c}(\textbf{r})$, and (f) the derived Galactic free--free emission $T_{\rm b}^{\rm Gff}(\textbf{r})$ at \SI{150}{\MHz}. All maps are shown in Cartesian projection (\num{8640} $\times$ \num{4320}) in equatorial coordinates on a logarithmic scale, except for the $T_{\rm e}(\textbf{r})$ and $\tau_{\rm c}(\textbf{r})$ that are shown on a linear scale. The $({\rm R.A., ~Dec.})$ = (\SI{0}{\degree}, \SI{0}{\degree}) is at the centre of each figure. The magenta asterisk marks the centre of region A analyzed in Section \ref{chap:results}, and the white and yellow asterisks show the centres of regions B and C (see Section \ref{chap:dis}) for comparison.} 
	\label{fig:maps}
\end{figure*}

\begin{figure}
	\centering
	\includegraphics[width=0.50\textwidth]{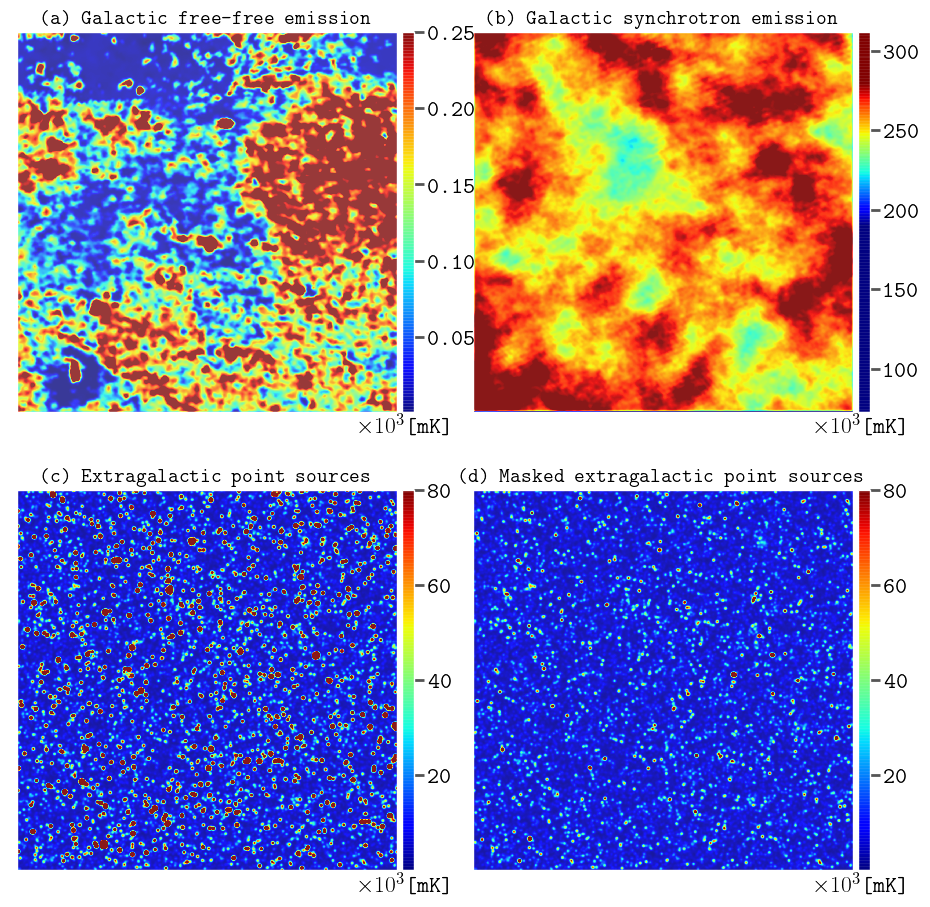}
	\caption{The foreground maps at \SI{150}{\MHz} of (a) Galactic free--free emission, (b) Galactic synchrotron emission, (c) Extragalactic point sources, and (d) Masked extragalactic point sources. Each map covers \SI[product-units=repeat]{10 x 10}{\degree} sky region and each color bar is in units of $\times$\SI{e3}{\mK}.}
	\label{fig:synsffps}
\end{figure}

We present the full-sky electron temperature $T_{\rm e}(\textbf{r})$ map proposed by \citet{Planck-X16}, the derived Galactic free--free optical depth $\tau_{\rm c}(\textbf{r})$ map, and the Galactic free--free emission $T_{\rm b}^{\rm Gff}(\textbf{r})$ in the bottom panels of Figure \ref{fig:maps} (the maps of $\tau_{\rm c}(\textbf{r})$ and $T_{\rm b}^{\rm Gff}(\textbf{r})$ at \SI{150}{\MHz} are shown for instance). We find that, at high latitudes, the electron temperature $T_{\rm e}(\textbf{r})$ is about \num{7000 \pm 1000} \si{\K}, and the optical depth is $\tau_{\rm c}(\textbf{r}) \lesssim \num{e-4}$ (optically thin). It is also found that the Galactic free--free emission is concentrated on the Galactic plane, which shows a typical brightness temperature of \SI{0.3}{\K} at high latitudes. In the Galactic plane, the Galactic free--free emission is optically thick due to the self-absorption, which becomes important at lower frequencies ($\sim$\SIrange{50}{200}{\MHz}), whereas at high latitudes this effect is presumably unimportant, i.e., the optically thin approximation holds. We present the \SI{150}{\MHz} example Galactic free--free emission map (\SI[product-units=repeat]{10 x 10}{\degree}) centred at ($\rm R.A.$, $\rm Dec.$) = (\SI{0}{\degree}, \SI{-30}{\degree}) (region A) in Figure \ref{fig:synsffps} (top left panel). The root-mean-square (rms) brightness temperatures of Galactic free--free emission at \numlist{120; 150; 190} \si{\MHz} are listed in Table \ref{Tab:rms-Tb}, respectively.

\begin{center}
	\begin{table}
	\caption{The rms brightness temperatures of the EoR signal and foreground components (Unit: \si{\mK}).}\label{Tab:rms-Tb}
		\resizebox{0.48\textwidth}{10mm}{ %
			\begin{tabular}{lcccccccccccccc}
				\hline
				\hline
				Components & \SI{120}{\MHz} & \SI{190}{\MHz}  & \SI{190}{\MHz}  \\
				\hline
				EoR signal  &  \num{22.5} & \num{12.4} & \num{3.30}\\
				Galactic free--free    &  \num{296} & \num{186} & \num{114}\\
				Galactic synchrotron  & \num{5.15e5}  & \num{2.61e5}  & \num{1.27e5}\\
				Extragalactic point sources  & \num{3.69e8} & \num{8.38e7} & \num{2.97e7} \\
				Masked extragalactic point sources  & \num{3.44e5} & \num{1.29e5} & \num{9.16e4} \\
				\hline
				\hline
			\end{tabular}
		}%
		\noindent \footnotesize{$^{}$ Note that the rms brightness temperatures are calculated over the sky coverage of \SI[product-units=repeat]{10 x 10}{\degree} on the \num{1800} $\times$ \num{1800} pixels.}\\
	\end{table}
    
\end{center} 

\subsection{Galactic Synchrotron Emission}
\label{chap:syn}

We construct the Galactic synchrotron map based on the Haslam \SI{408}{\MHz} map \citep{Haslam81,Haslam82}. Specifically, we utilize the high-resolution Haslam \SI{408}{\MHz} map\footnote{The reprocessed Haslam \SI{408}{\MHz} map: \url{http://www.jb.man.ac.uk/research/cosmos/haslam_map/}} reprocessed by \citet{Remazeilles15} ($N_{\rm side}$ = \num{2048}, pixel size is \SI{\sim 1.72}{\arcminute}; RH408 hereafter), which exhibits better calibration and extragalactic source removal. The Galactic synchrotron sky map at required frequency is then obtained by extrapolating the RH408 \si{\MHz} map $T_{\rm RH408}^{\rm {Gsyn}}(\textbf{r})$ using a power-law form (e.g., \citealt{Wang10,Thorne17,Li19})

\begin{equation}
\label{equ:syn}
T_{\rm b}^{\rm {Gsyn}}(\textbf{r}) = T_{\rm RH408}^{\rm {Gsyn}}(\textbf{r})~~(\frac{\nu}{\rm RH408})^{-\alpha^{\rm Gsyn}(\textbf{r})},
\end{equation} 

\noindent where $\alpha^{\rm Gsyn}(\textbf{r})$ is the corresponding spectral index map. Although \citet{Lawson87} (L87 hereafter) proposed a spectral index map within \SIrange{38}{1420}{\MHz}, there is no data of the southern sky. To account for the region-to-region index variations, the all-sky synchrotron spectral index map provided by \citet{Miville08} (MD08 hereafter) is employed. Compared with the L87 map ($\simeq$ \numrange{2.4}{2.7}), the MD08 map shows values of $\simeq$ \numrange{2.7}{3.3}, which may cause the synchrotron emission to be about \num{30}\% brighter than the actual situation. Whereas the Galactic synchrotron emission is used just for comparison, which does not affect the analyses of Galactic free--free emission in this work. Same as the Galactic free--free emission, the self-absorption of Galactic synchrotron becomes important in the optically thick region, especially in the Galactic plane \citep{Zheng12}. We show the \SI{150}{\MHz} example Galactic synchrotron emission map (\SI[product-units=repeat]{10 x 10}{\degree}) centred at ($\rm R.A.$, $\rm Dec.$) = (\SI{0}{\degree}, \SI{-30}{\degree}) in Figure \ref{fig:synsffps} (top right panel). The corresponding rms brightness temperatures of Galactic synchrotron emission at \numlist{120; 150; 190} \si{\MHz} are listed in Table \ref{Tab:rms-Tb}, respectively. 

\subsection{Extragalactic Point Sources}
\label{chap:ptr}

Following our previous works \citep{Wang10,Wang13,Li19} five types of extragalactic sources are simulated, including (1) normal star-forming and starburst galaxies, (2) radio-quiet active galactic nucleus (AGNs), (3) Fanaroff-Riley type I (FRI) and type II (FRII) AGNs, (4) GHz-peaked spectrum (GPS) AGNs, and (5) compact steep spectrum (CSS) AGNs. For the former three types we adopt the flux densities, spatial structures, spectra, and angular clusterings proposed by \citet{Wilman08}, while for the latter two types we employ the quantities by applying their corresponding luminosity functions and spectral models \citep{Wang10}.

The peeling strategy is employed for the brightest point sources (masked extragalactic point sources, hereafter; e.g., \citealt{Mitchell08,Intema09,Mort17}), and we assume the flux  $S_{150} > \SI{54}{\mJy}$ is removed to keep the relative appropriate dynamic range by following the previous works (e.g., \citealt{Liu09,Pindor11,Li19}). The masked extragalactic point sources emission at \SI{150}{\MHz} is also shown in Figure \ref{fig:maps} (bottom right panel), compared to the raw extragalactic point sources emission (the bottom left panel of Figure \ref{fig:maps}), the rms brightness temperatures of point sources are significantly reduced from (\numlist{36.9; 8.38; 2.97}) $\times$ \SI{e7}{\mK} to (\numlist{34.37; 12.88; 9.16}) $\times$ \SI{e4}{\mK} at \numlist{120; 150; 190} \si{\MHz}, respectively (see Table \ref{Tab:rms-Tb}).


\section{Simulation of EoR Signal}
\label{chap:21cm}

\begin{figure*}
	\centering
	\includegraphics[width=0.49\textwidth]{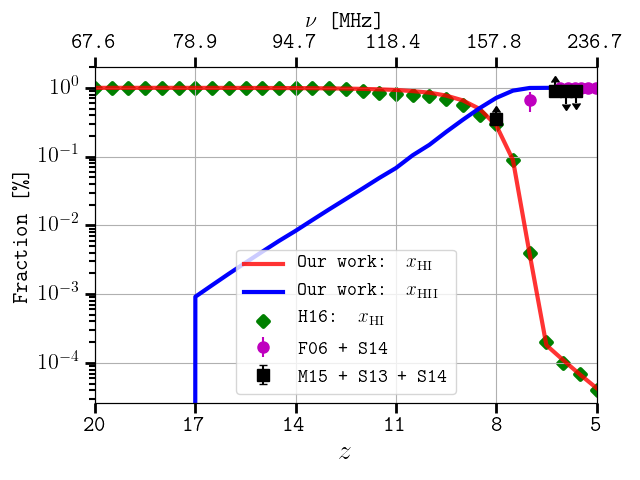}
	\includegraphics[width=0.49\textwidth]{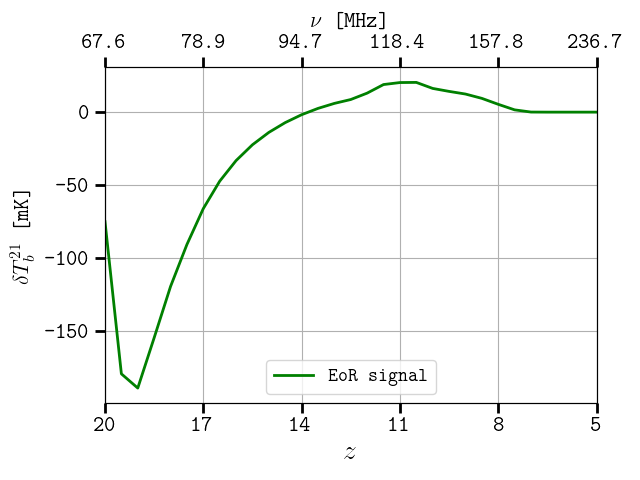}
	\caption{The comparison of the volume-weighted average neutral hydrogen fraction between our simulation (red solid line) and H16 (green diamond marker), with the corresponding comparison of the ionized hydrogen fraction between our simulation (blue solid line) and the observed constraints (the purple points and the black squares) proposed by F06, S13, S14, and M15 (left panel). The right panel shows the brightness temperatures of EoR signal (green solid line) within \SIrange{67.6}{236.7}{\MHz} ($z$ = \numrange{20}{5}).}
	\label{fig:t21-xHII}
\end{figure*}

We simulate the brightness temperatures of the 21-cm signal during the EoR using the semi-numerical {\sc Simfast21}\footnote{\url{https://github.com/mariogrs/Simfast21}} code (\citealt{Santos10,Hassan16}; H16 hereafter). The {\sc Simfast21} generates a series of three-dimensional (3D) cubes of evolved matter density, ionization, peculiar velocity, and spin temperature fields that can be used to compute the brightness temperature of the EoR signal. This code employs approximate methods to realize the physical processes (e.g., Ly$\alpha$, X-ray, star formation ratio, etc) and shows a good agreement with the fully numerical simulation. Our simulation evolves from the initial redshift $z_{\rm i}$ = \num{100} to the final redshift $z_{\rm f}$ = \num{5} on a $1024^3$ box with physical dimensions of $1.6^3$ comoving $\rm Gpc^{3}$, which corresponds to a field of $\theta_x$ = $\theta_y$ $\approx$ \SI{9.88}{\degree}, a pixel resolution of $\Delta\theta_x$ = $\Delta\theta_y$ $\approx$ \SI{0.58}{\arcminute}, and a frequency depth of $\Delta\nu$ $\approx$ \SI{92.95}{\MHz}. We compare the volume-weighted average neutral hydrogen fraction and the corresponding ionized hydrogen fraction with the H16 simulation and some observed constraints (e.g., \citealt{Fan06,Schroeder13,Schenker14,McGreer15}) in Figure \ref{fig:t21-xHII} (left panel), and find that our simulation shows the consistent result with the H16 simulation and the observations. In Figure \ref{fig:t21-xHII} (right panel) we show the derived brightness temperatures of EoR signal $\delta T_{b}^{21}$ in the \SIrange{67.6}{236.7}{\MHz} ($z$ = \numrange{20}{5}) frequency band.   

\begin{figure*}
	\centering
	\includegraphics[width=1.0\textwidth]{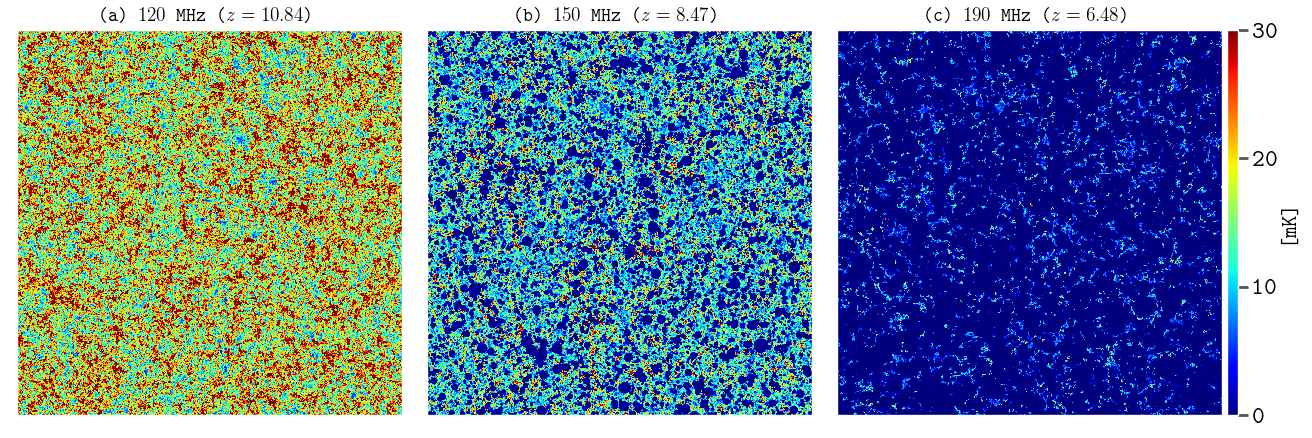}
	\caption{The simulated brightness temperatures of the EoR signal at \num{120}, \num{150}, and \num{190} \si{\MHz} ($z$ = \numlist{10.84; 8.47; 6.48}). The sky region coverage is \SI[product-units=repeat]{10 x 10}{\degree} and the color bar is in units of \si{\mK}.}
	\label{fig:21cm}
\end{figure*}

The outputs of {\sc Simfast21} are the 21-cm brightness temperature cubes, which are known as the ``coeval cubes'' that cannot be directly observed. Following the method outlined in \citet{Mellema06} and \citet{Datta12}, we create the observable ``light-core'' object of the EoR signal based on these ``coeval cubes''. From the derived ``light-core'' object, we extract three subsets with $51$-channels (a channel width of \SI{160}{\kHz}), and construct our final tiled data cube with dimensions of ($\theta_x$, $\theta_y$, $\Delta\nu$) = (\SI{10}{\degree}, \SI{10}{\degree}, \SI{8}{\MHz}). We present the simulated sky maps of the EoR signal $\delta T_{\rm b}^{21}$ at \num{120} ($z$ = \num{10.84}; left panel), \num{150} ($z$ = \num{8.47}; middle panel), and \SI{190}{\MHz} ($z$ = \num{6.48}; right panel) in Figure \ref{fig:21cm}, and list the rms brightness temperatures of $\delta T_{\rm b}^{21}$ in Table \ref{Tab:rms-Tb}.

\section{Simulation of SKA Observation}
\label{chap:sim-obs}

To incorporate the instrumental effects (i.e., the instrumental noise and the beam) of radio interferometers, we have employed the latest SKA1-Low layout configuration%
\footnote{\raggedright%
	SKA1-Low Configuration Coordinates:
	\url{https://astronomers.skatelescope.org/wp-content/uploads/2016/09/SKA-TEL-SKO-0000422_02_SKA1_LowConfigurationCoordinates-1.pdf} (released on 2016 May 31)} to simulate the SKA ``observed'' images. The SKA1-Low interferometer layout includes \num{512} stations, with \num{224} stations randomly distributing within the ``core'' region (\SI{1000}{\meter} in diameter), and others scattering in ``clusters'' regions, which form \num{3} spiral arms up to a radius of \SI{\sim 35}{\kilo\meter}. Each station includes \num{256} antennas, which are randomly distributed in a circular region of \SI{35}{\meter} in diameter with a minimum separation of $d_{\rm min}$ = \SI{1.5}{\meter} (e.g., \citealt{Mort17}).

We use the {\sc {OSKAR}}\footnote{\raggedright OSKAR: \url{https://github.com/OxfordSKA/OSKAR} (version 2.7.0)} \citep{Mort10} simulator to perform SKA observations for \SI{6}{\hour} to obtain the visibility data based on both the foreground and EoR images (see Section \ref{chap:foregrounds} and Section \ref{chap:21cm}). The simulated visibility data are then imaged through the {\sc {WSClean}}\footnote{%
WSClean: \url{https://sourceforge.net/p/wsclean} (v2.6)} software \citep{Offringa14} using Briggs weighting with zero robustness to consider both the noise level and the spatial resolution \citep{Briggs95,Thompson17,Li19}, and the created images are cropped to keep only the central regions to avoid the insufficient CLEAN problem in the marginal regions. To be specific, for frequency bands of \numrange{116}{124}, \numrange{146}{154}, and \SIrange{186}{194}{\MHz} we choose to keep the central \SI[product-units=repeat]{6 x 6}{\degree}, \SI[product-units=repeat]{5 x 5}{\degree}, and \SI[product-units=repeat]{4 x 4}{\degree} regions, respectively, since the telescope's field of view (FoV) is inversely proportional to the frequency.

We apply the CLEAN algorithm with joined-channel deconvolution \citep{Offringa17} to create the foreground cubes in each frequency band. For the EoR signal, however, we directly use the dirty image to create the EoR image because the CLEAN algorithm does not work well for extremely faint emission. Hence we obtain the SKA ``observed'' image cubes of EoR signal, Galactic synchrotron emission, Galactic free--free emission, and the masked extragalactic point sources in the \numrange{116}{124}, \numrange{146}{154}, and \SIrange{186}{194}{\MHz} frequency bands, which will be used to carry out the power spectrum analyses.

\section{Power Spectra and EoR Window}
\label{chap:ps-eor}

The redshifted 21-cm signals observed at different frequencies are expected to form a 3D data cube, with two spatial dimensions describing the transverse distances across the sky and the one frequency dimension depicting the line-of-sight distance. For each foreground emission cube, the two angular dimensions describe the same sky coverage as the EoR signal, but the one frequency dimension depicts the emission distribution in the frequency space (i.e., spectrum), which is different from the EoR signal. The 3D power spectrum $P(k_x, k_y, k_z)$ is calculated from each image cube, which is spherically symmetric within a limited redshift range ($\Delta z \sim 0.5$, when an \SI{8}{\MHz} frequency bandwidth is adopted). The Blackman-Nuttall window function is applied to the frequency dimension before calculating the $P(k_x, k_y, k_z)$ to suppress the significant side-lobes in the Fourier transform \citep{Trott15,Chapman16,Li19}. 
The corresponding one-dimensional (1D) power spectrum $P(k)$ is calculated by averaging the $P(k_x, k_y, k_z)$ in spherical shells of radii $k$, which achieves a relatively higher signal-to-noise ratio than the direct imaging observations (e.g., \citealt{Morales04,Datta10}). The dimensionless variant of the 1D power spectrum $\Delta^{2}(k)$ = $P(k)k^{3}/(2\pi^{2})$ is employed in our work, as commonly adopted in the literature (e.g., \citealt{Li19}).
 
We further calculate the corresponding two-dimensional (2D) power spectrum $P(k_{\perp}, k_{||})$ by averaging the 3D power spectrum $P (k_x, k_y, k_z)$ over the angular annuli of radii $k_{\perp} \equiv \sqrt{k_{x}^{2} + k_{y}^{2}}$ for each line-of-sight plane $k_{||} \equiv k_z$. It is found that in the ($k_{\perp}, k_{||}$) plane the spectrally-smooth Galactic free--free emission dominates the low-$k_{||}$ region, but some purely angular ($k_{\perp}$) modes of the foreground signals can be thrown into the line-of-sight ($k_{||}$) dimension (called mode mixing), due to the complicated instrumental and observational effects (e.g., chromatic primary beams, calibration errors). Consequently, an expanded wedge-like contamination region appears at the bottom right in the ($k_{\perp}, k_{||}$) plane, which is known as the foreground wedge \citep{Datta10,Morales12,Liu14}. The top left corner in the ($k_{\perp}, k_{||}$) plane, on the other hand, is almost free from the foreground contamination, namely the EoR window, whose description was proposed by \citet{Thyagarajan13}

\begin{equation}
\label{equ:EoR}
k_{||} \geq \frac{H(z)D_{\rm M}(z)}{(1+z)c} [k_{\perp}~{\rm {sin} \Theta} + \frac{2\pi w f_{21}}{(1 + z) D_{\rm M}(z) B}],
\end{equation}

\noindent where $H(z)$ is the Hubble parameter at redshift $z$, $D_{\rm M}(z)$ is the transverse comoving distance, $B$ = \SI{8}{MHz} is the frequency bandwidth of the image cube, $w$ ($\propto B$) is the number of characteristic convolution widths for the spillover region caused by the variations in instrumental frequency response, $\Theta$ is the angular distance of the foreground sources from the field centre, and $f_{21}$ = \SI{1420.4}{\MHz} is the rest frequency of the 21-cm emission line.

\section{Results}
\label{chap:results}

To quantitatively evaluate the contamination, we calculate the 1D power spectra and compare the power of Galactic free--free emission with that of the EoR signal. Meanwhile, we calculate the 2D power spectra and carry out the comparison between the Galactic free--free emission and the EoR signal inside the EoR window. 

\subsection{1D Power Spectra}
\label{chap:results-ps1d}

\begin{figure*}
	\centering
	\includegraphics[width=1.0\textwidth]{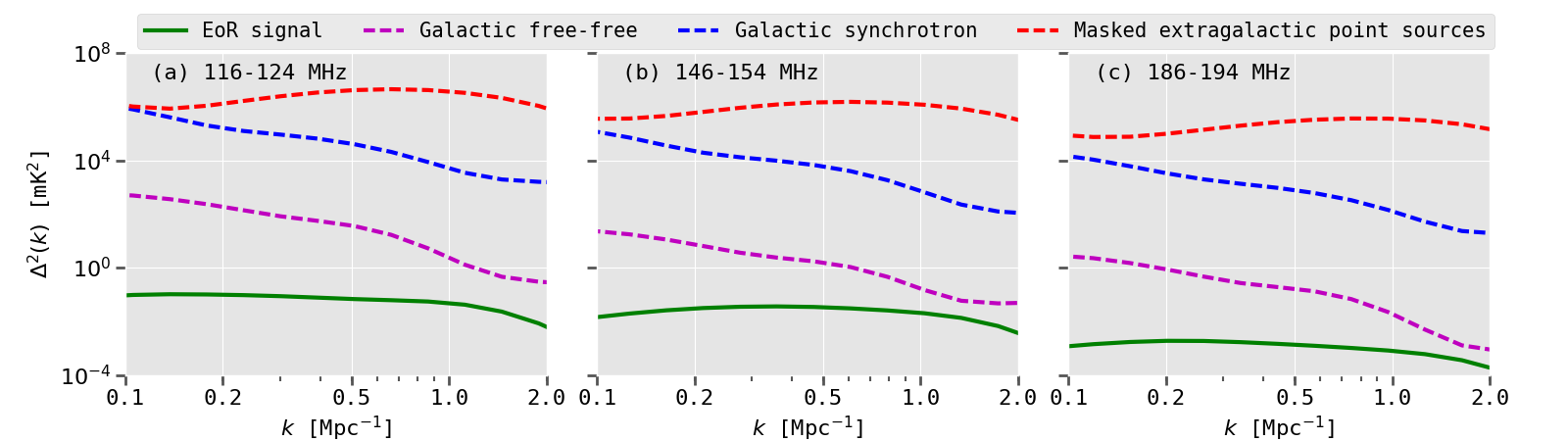}
	\caption{The 1D power spectra $\Delta^{2}(k)$ of the EoR signal (green solid line), Galactic free--free emission (magenta dashed line), Galactic synchrotron emission (blue dashed line), and masked extragalactic point sources (red dashed line) in the (a) \SIrange{116}{124}{\MHz} (left panel), (b) \SIrange{146}{154}{\MHz} (middle panel), and (c) \SIrange{186}{194}{\MHz} (right panel) frequency bands.}
	\label{fig:ps1d-comp}
\end{figure*}

We calculate the 1D dimensionless power spectra $\Delta^{2}(k)$ of EoR signal and foreground components from the image cubes obtained in Section \ref{chap:sim-obs}. The comparisons of the 1D power spectra $\Delta^{2}(k)$ between the Galactic free--free emission and the EoR signal in the \numrange{116}{124}, \numrange{146}{154}, and \SIrange{186}{194}{\MHz} frequency bands are shown in Figure \ref{fig:ps1d-comp}, in which we also present the $\Delta^{2}(k)$ of the Galactic synchrotron emission and the masked extragalactic point sources (i.e., the case with the brightest point sources removed) for comparison. It is obvious that the contamination caused by the Galactic free--free emission is a function of position in the $k$-space, and the power spectra show that the Galactic free--free emission is more luminous than the EoR signal by about \numrange{e3.5}{e2.0}, \numrange{e3.0}{e1.3}, and \numrange{e2.5}{e1.0} times on scales of $\SI{0.1}{\per\Mpc} < k < \SI{2}{\per\Mpc}$ in the \numrange{116}{124}, \numrange{146}{154}, and \SIrange{186}{194}{\MHz} frequency bands, respectively. These results show that the Galactic free--free emission is a severe foreground contaminating component and should be accurately removed in future low-frequency radio experiments, such as the international SKA project. Indeed, we can see clearly that the Galactic synchrotron emission and the masked extragalactic point sources are the main contaminating sources on scales of $\SI{0.1}{\per\Mpc} < k < \SI{2}{\per\Mpc}$ in three frequency bands, whose power are more luminous than that of the EoR signal by about $4$ to $6$ orders of magnitude. 

\subsection{2D Power Spectra}
\label{chap:results-ps2d}

\begin{figure}
	\centering
	\includegraphics[width=0.5\textwidth]{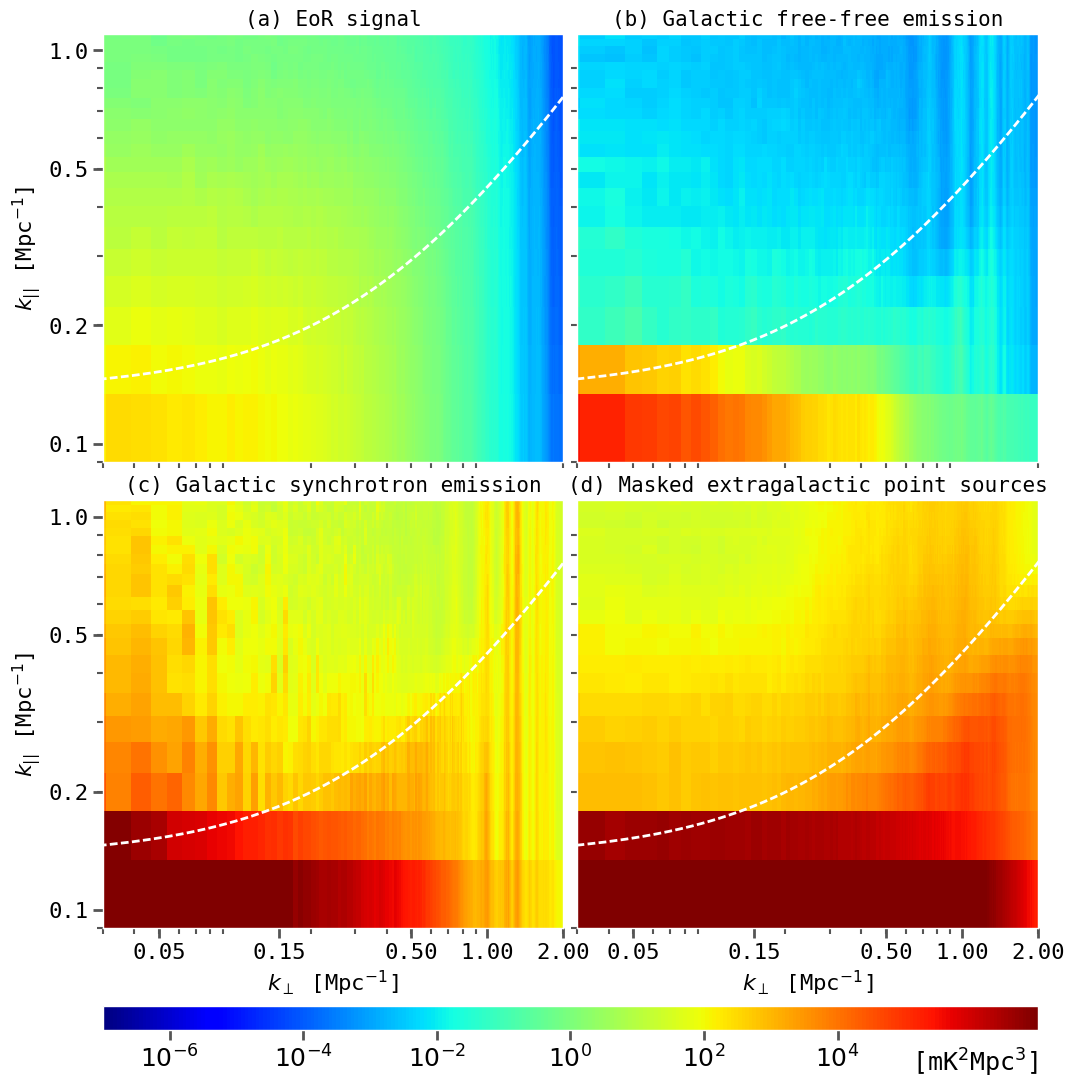}
	\caption{The \SIrange{146}{154}{\MHz} 2D power spectra $P(k_{\perp}, k_{||})$ of (a) the EoR signal, (b) Galactic free--free emission, (c) Galactic synchrotron emission, and (d) masked extragalactic point sources. The white dashed lines mark the boundary between the EoR window (top left) and the foreground wedge (bottom right). All panels share the same logarithmic scale in units of $\rm {mK^{2}Mpc^{3}}$.}
	\label{fig:ps2d}
\end{figure}

The 2D power spectra $P(k_{\perp}, k_{||})$ of the EoR signal, Galactic free--free emission, Galactic synchrotron emission, and the masked extragalactic point sources are shown in Figure \ref{fig:ps2d} (take \SIrange{146}{154}{\MHz} frequency band for instance). We find that the EoR signal distributes its power across all $k_{||}$ modes, which illustrates its rapid fluctuations along the line-of-sight dimension, while the spectrally-smooth foreground components dominate the low-$k_{||}$ ($k_{||}$ $\lesssim$ \SI{0.2}{\per\Mpc}) regions. With regard to the angular dimension ($k_{\perp}$), the Galactic free--free emission and Galactic synchrotron emission dominate the power on scales of $k_{\perp}$ $\lesssim$ \SI{0.5}{\per\Mpc}, while the masked extragalactic point sources distribute the power on scales of $k_{\perp}$ $\lesssim$ \SI{1.0}{\per\Mpc}. 

\begin{figure*}
	\centering
	\includegraphics[width=1.0\textwidth]{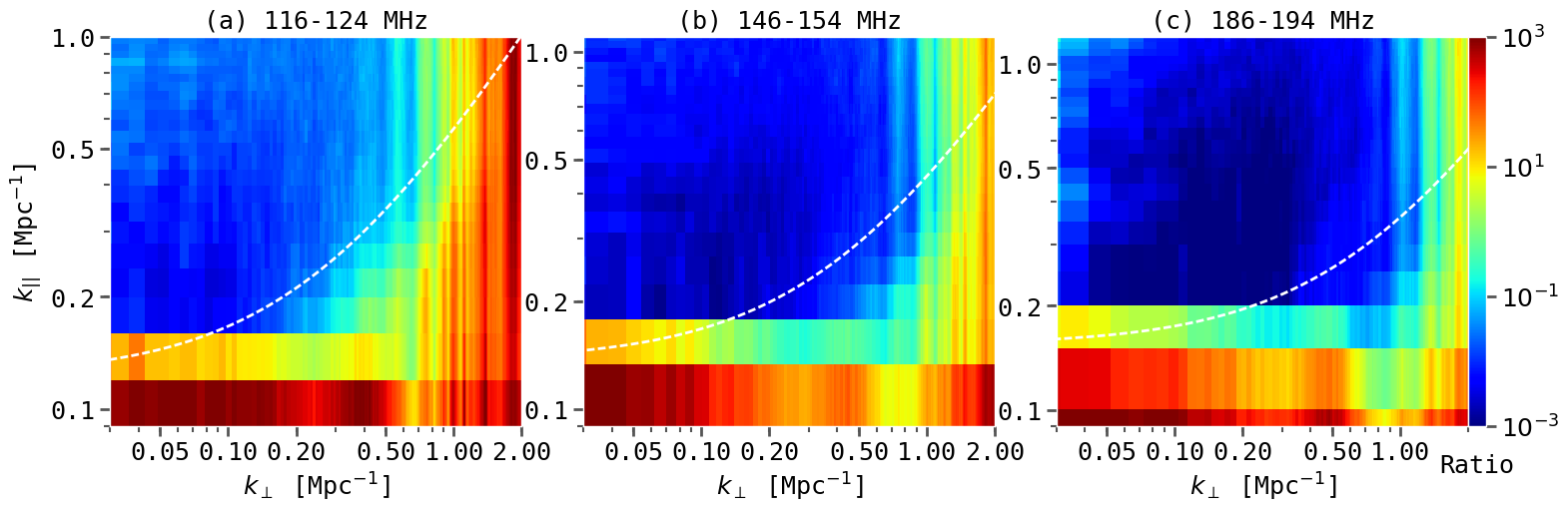}
	\caption{The 2D power spectra ratios $R(k_{\perp}, k_{||})$ of Galactic free--free emission to the EoR signal in the (a) \SIrange{116}{124}{\MHz} (left panel), (b) \SIrange{146}{154}{\MHz} (middle panel), and (c) \SIrange{186}{194}{\MHz} (right panel) frequency bands. The white dashed lines mark the EoR window boundary, and the EoR window is above the white line. All panels share the same color bar on a logarithmic scale.}
	\label{fig:ps2d-ratio}
\end{figure*}

To further quantify the contamination caused by Galactic free--free emission when the foreground avoidance method is adopted, we define an EoR window in the ($k_{\perp}, k_{||}$) plane according to Equation \ref{equ:EoR} with a $w = 3$ configuration and the SKA1-Low's FoV (i.e., $\Theta$ = \SI{6}{\degree}, \SI{5}{\degree}, and \SI{4}{\degree} in the \numrange{116}{124}, \numrange{146}{154}, and \SIrange{186}{194}{\MHz} frequency bands, respectively) to avoid the heavily contaminating foreground wedge (Figure \ref{fig:ps2d} and Figure \ref{fig:ps2d-ratio}). We find that \num{\sim 50}\% power of the EoR signal and \num{\sim 95}\% power of the Galactic free--free emission are lost in the foreground wedge. We calculate the 2D power spectrum (amplitude squared) ratio $R(k_{\perp}, k_{||})$, which is defined as $R(k_{\perp}, k_{||})$ = $P_{\rm Gff}(k_{\perp}, k_{||})$ / $P_{\rm 21cm}(k_{\perp}, k_{||})$, where $P_{\rm Gff}(k_{\perp}, k_{||})$ and $P_{\rm 21cm}(k_{\perp}, k_{||})$ are the 2D power spectra of Galactic free--free emission and the EoR signal, respectively. As illustrated in Figure \ref{fig:ps2d-ratio}, we find that the $R(k_{\perp}, k_{||})$ $\ll$ 1 for much of the region inside the EoR window, i.e. the contamination imposed by Galactic free--free emission on the EoR signal can be ignored. However, the power leaked by Galactic free--free emission can still be significant, for the $R(k_{\perp}, k_{||})$ is greater than 1 on angular scales of $k_{\perp}$ $\gtrsim$ \SI{0.5}{\per\Mpc}, $k_{\perp}$ $\gtrsim$ \SI{0.8}{\per\Mpc}, and $k_{\perp}$ $\gtrsim$ \SI{1.0}{\per\Mpc} in the \numrange{116}{124}, \numrange{146}{154}, and \SIrange{186}{194}{\MHz} frequency bands, respectively. It is also shown that Galactic free--free emission causes more serious contamination toward lower frequencies (\SI{\sim 116}{\MHz}) because of its relatively steep spectral index ($\sim -2.1$).

\begin{figure}
	\centering
	\includegraphics[width=0.50\textwidth]{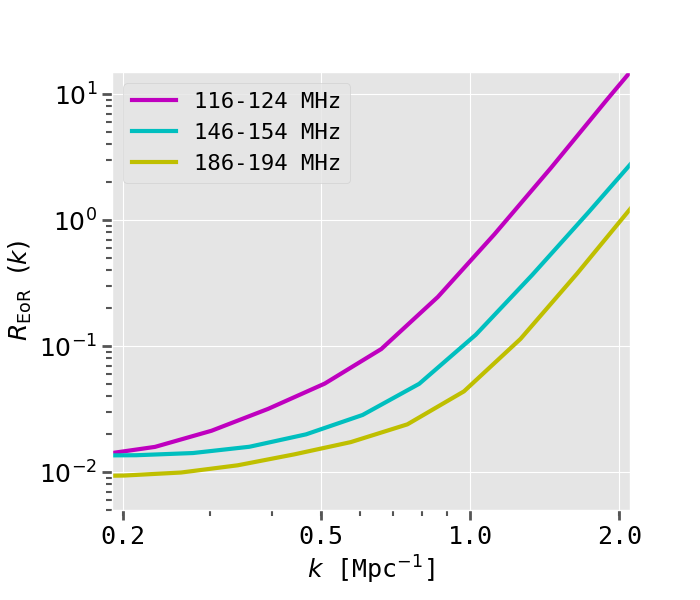}
	\caption{The 1D power ratios $R_{\rm EoR}(k)$ inside the EoR window of Galactic free--free emission to the EoR signal. The magenta, cyan, and yellow solid lines show the ratios in \numrange{116}{124}, \numrange{146}{154}, and \SIrange{186}{194}{\MHz} frequency bands, respectively.}
	\label{fig:ps1d-ratio}
\end{figure}

To better constrain the contamination caused by the Galactic free--free emission, by averaging the modes only inside the above properly defined EoR window, we calculate the 1D power spectrum ratios $R_{\rm EoR}(k)$ of Galactic free--free emission to the EoR signal and present the results in Figure \ref{fig:ps1d-ratio}. We find that, compared to Figure \ref{fig:ps1d-comp}, the 1D power ratio is suppressed by about $3$ orders of magnitude, which demonstrates that the EoR window is a powerful tool for avoiding the strong foreground, as the $R_{\rm EoR}(k)$ on scales of \SI{0.5}{\per\Mpc} are only about \num{6}\%, \num{2}\%, and \num{1}\% in the \numrange{116}{124}, \numrange{146}{154}, and \SIrange{186}{194}{\MHz} frequency bands, respectively. However, the power of Galactic free--free emission leaked into the EoR window can not be ignored, since the $R_{\rm EoR}(k)$ can be up to about \num{200}\%, \num{60}\%, and \num{15}\% on scales of \SI{1.2}{\per\Mpc} in three frequency bands, respectively. Based on the above calculations, we conclude that the Galactic free--free emission is a non-negligible contaminating foreground component for EoR observations. Even within the EoR window where most of the strong foreground contamination is avoided, the EoR signal can still be contaminated by the Galactic free--free emission, especially toward lower frequencies (\SI{\sim 116}{\MHz}). 

\section{Discussion}
\label{chap:dis}

\subsection{Impacts of Sky Positions}

\begin{figure}
	\centering
	\includegraphics[width=0.50\textwidth]{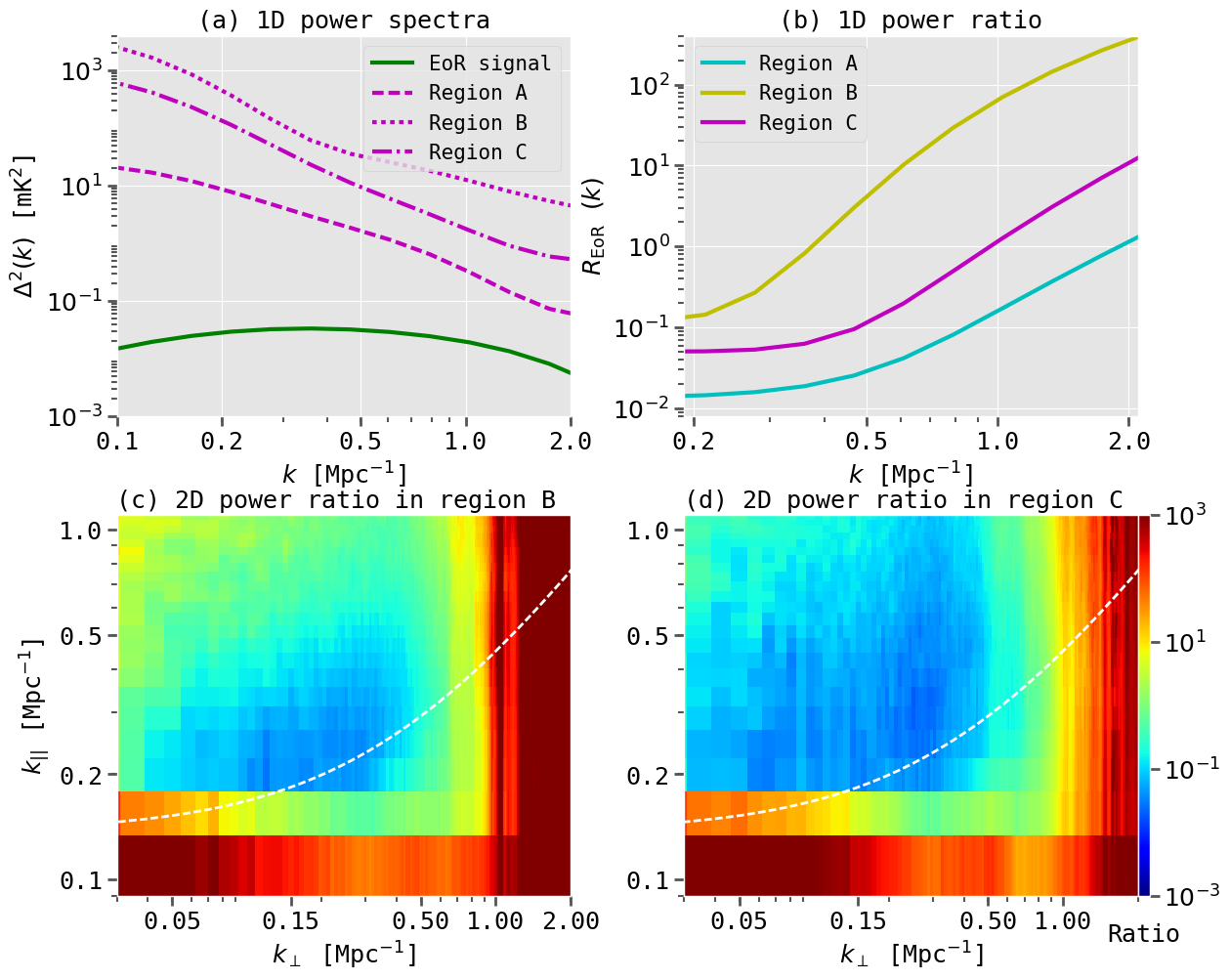}
	\caption{(a) The \SIrange{146}{154}{\MHz} 1D power spectra $\Delta^{2}(k)$ of Galactic free--free emission in regions A, B, and C (magenta dashed, dotted, and dashed-dotted lines, respectively), and the EoR signal (green solid line). (b) The corresponding 1D power ratios inside the EoR window $R_{\rm {EoR}}(k)$ of Galactic free--free emission to the EoR signal in regions A, B, and C (cyan, yellow, and magenta solid lines, respectively). (c) The corresponding 2D power spectra ratios $R(k_{\perp}, k_{||})$ of Galactic free--free emission to the EoR signal in region B. (d) The corresponding $R(k_{\perp}, k_{||})$ in region C. The lower two panels are shown on a logarithmic scale, and the white dashed lines mark the EoR window boundaries.}
	\label{fig:3regions}
\end{figure}

It is generally known that the above analyses will vary region-to-region and the foreground at lower latitudes will impose the severer contamination on the EoR signal. To measure the impact of sky positions, following the work of \citet{Sims16}, we choose two other (\SI[product-units=repeat]{10 x 10}{\degree}) sky regions centred at $({\rm R.A., ~Dec.})$ = (\SI{50}{\degree}, \SI{-30}{\degree}) corresponding to $({l,~ b})$ = (\SI{227}{\degree}, \SI{-57}{\degree}) (region B) and $({\rm R.A.,~ Dec.})$ = (\SI{310}{\degree}, \SI{-30}{\degree}) corresponding to $({l, ~b})$ = (\SI{14}{\degree}, \SI{-35}{\degree}) (region C). The regions B and C show the same declinations (different latitudes) as region A and are preferred to perform SKA simulation. The rms brightness temperatures of Galactic free--free emission at \SI{150}{\MHz} for regions B and C are \SI{619}{\mK} and \SI{411}{\mK}, respectively. The \SIrange{146}{154}{\MHz} 1D power spectra $\Delta^{2}(k)$ of Galactic free--free emission in regions B and C are calculated and compared with the $\Delta^{2}(k)$ in region A centred at $({\rm R.A.,~ Dec.})$ = (\SI{0}{\degree}, \SI{-30}{\degree}) in Figure \ref{fig:3regions} (top left panel). We find that the 1D power spectra $\Delta^{2}(k)$ of regions B and C are about \num{e1.5} and \num{e1.0} times more luminous than that of region A on scales of $\SI{0.1}{\per\Mpc} < k < \SI{2}{\per\Mpc}$, respectively. In addition, for regions B and C, we calculate the 1D power ratios $R_{\rm EoR}($k$)$ of Galactic free--free emission to the EoR signal by averaging the modes only inside the properly defined EoR window and show the results in Figure \ref{fig:3regions} (top right panel). Compared to the result of region A (cyan solid line), we find that the Galactic free--free emissions in regions B and C cause severer contamination on the EoR signal, for the $R_{\rm EoR}($k$)$ there can be up to about \num{500}\%--\num{12000}\% and \num{20}\%--\num{400}\% on scales of $\SI{0.5}{\per\Mpc} \lesssim k \lesssim \SI{1.2}{\per\Mpc}$ in the \SIrange{146}{154}{\MHz} frequency band, respectively. We further calculate the \SIrange{146}{154}{\MHz} 2D power spectra ratios $R(k_{\perp}, k_{||})$ of Galactic free--free emission to the EoR signal in regions B and C and present the results in Figure \ref{fig:3regions} (bottom panels). It is clearly shown that, compared to the $R(k_{\perp}, k_{||})$ in region A, the $R(k_{\perp}, k_{||}) \gg 1$ for most modes in regions B and C, especially on scales of $k_{\perp}$ $\gtrsim$ \SI{0.5}{\per\Mpc}, meaning that the Galactic free--free emissions in regions B and C have a greater impact on the EoR detection, which is consistent with the analyses of the 1D power spectra (top left panel) and the 1D power ratios inside the EoR window (top right panel). 

\subsection{Impacts of Frequency Artifacts}

\begin{figure*}
	\centering
	\includegraphics[width=1.0\textwidth]{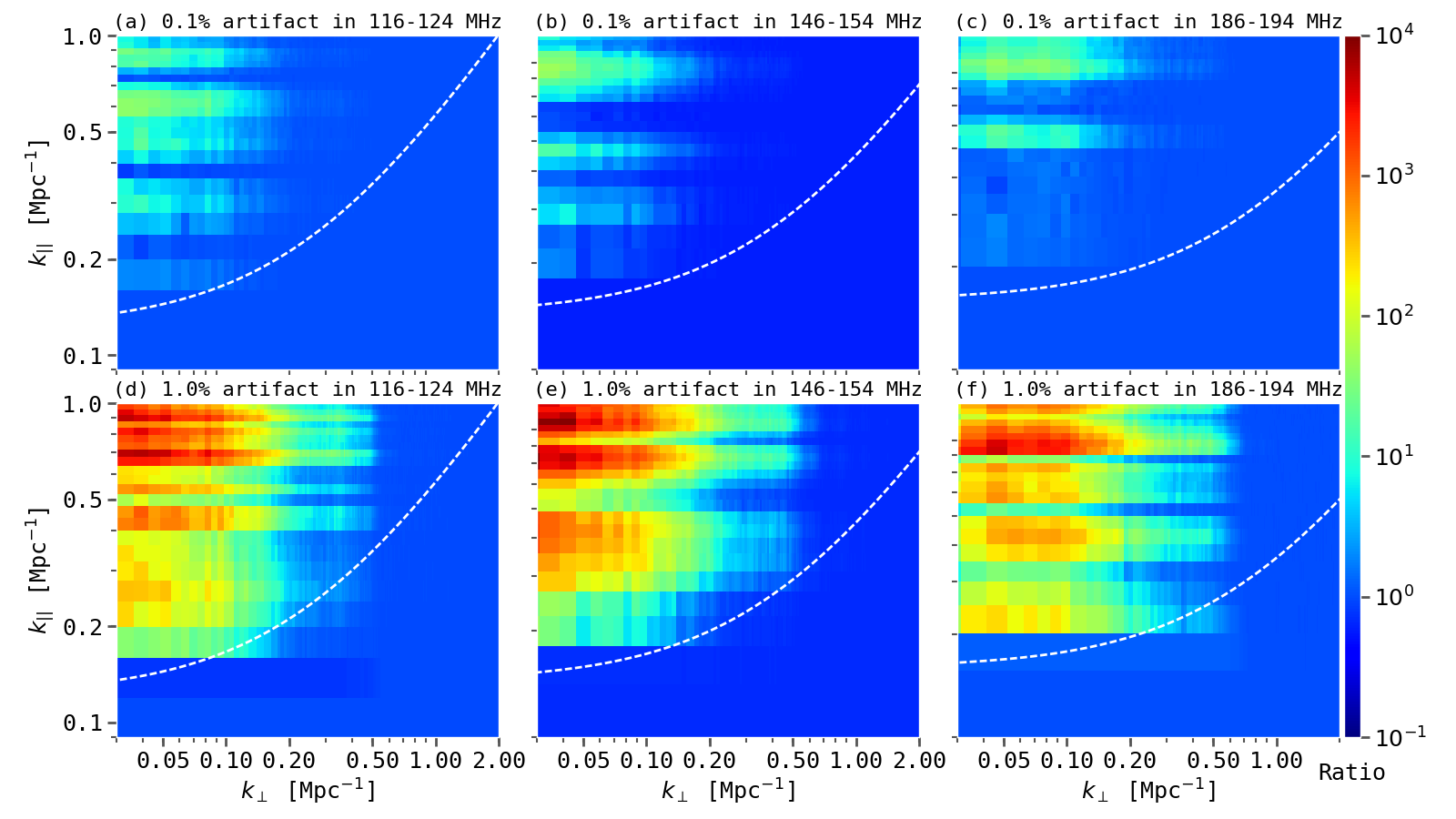} 
	\caption{The 2D power spectrum ratios $R_{\rm arti}(k_{\perp}, k_{||})$ of Galactic free--free emissions that are derived from the artifact image cubes and the original ones. The upper and lower rows show the $R_{\rm arti}(k_{\perp}, k_{||})$ when frequency artifacts are $A_{\rm arti} = 0.1\%$ and $A_{\rm arti} = 1\%$, respectively. The left, middle, and right columns show the $R_{\rm arti}(k_{\perp}, k_{||})$ in the \numrange{116}{124}, \numrange{146}{154}, and \SIrange{186}{194}{\MHz} frequency bands, respectively. The dashed white lines mark the EoR window boundaries. All panels share the same color bar on a logarithmic scale.}
	\label{fig:sff-artifact}
\end{figure*}

The obtained image cubes may exist the frequency artifacts due to miscalibration and various instrumental effects. To effectively simulate the bandpass gain errors (i.e., miscalibration of the bandpass), we multiply each image of the 3D image cube by a random number generated from a Gaussian distribution with unity mean \citep{Chapman16,Li19}. We then evaluate the impact of the frequency artifacts on the power spectrum by comparing the power spectrum calculated from the modified image cube with that derived from the original case. The residual calibration error in the frequency channels has been proved to be about 0.1\%--1\% (e.g., \citealt{Barry16,Ewall17}), and two extreme cases are investigated, i.e., frequency artifacts of amplitude $A_{\rm arti}$ = 0.1\% and $A_{\rm arti}$ = 1\% are realized by setting $\sigma = 0.001$ and $\sigma = 0.01$ for their Gaussian distributions, respectively. 

To estimate the impact of frequency artifacts on the Galactic free--free emission, we calculate the 2D power spectrum ratio $R_{\rm arti}(k_{\perp}, k_{||})$ of the artifact image cube to the original case (Section \ref{chap:results}), and present the $R_{\rm arti}(k_{\perp}, k_{||})$ with either $A_{\rm arti} = 0.1\% $ or $1\%$ in the \numrange{116}{124}, \numrange{146}{154}, and \SIrange{186}{194}{\MHz} frequency bands in Figure \ref{fig:sff-artifact}. We find that the 2D power spectra of Galactic free--free emission are seriously damaged by the frequency artifacts, i.e., on scales of $k_{\perp}$ $\lesssim$ \SI{0.2}{\per\Mpc} and $k_{||}$ $\gtrsim$ \SI{0.2}{\per\Mpc}, adding $A_{\rm arti} = 0.1\%$ causes the power of Galactic free--free emission to be about $4$, $3$, and $2$ times stronger in the \numrange{116}{124}, \numrange{146}{154}, and \SIrange{186}{194}{\MHz} frequency bands, respectively, and the corresponding increasing powers are about $400$, $300$, and $200$ times for $A_{\rm arti} = 1\%$. Meanwhile, we test the impact of frequency artifacts on other foreground components and find that the power of Galactic synchrotron emission becomes about $2.3$, $1.3$, and $1.1$ times stronger for $A_{\rm arti} = 0.1\%$ and about $100$, $40$, and $30$ times stronger for $A_{\rm arti} = 1\%$, and the power of the masked extragalactic point sources will be about $6$, $4$, and $3$ times stronger for $A_{\rm arti} = 0.1\%$ and about $800$, $600$, and $400$ times stronger for $A_{\rm arti} = 1\%$ on the same scales in three frequency bands, respectively. We also evaluate the same effects on the EoR signal by adding $A_{\rm arti} = 0.1\%$ and $1\%$ to the EoR image cube, but find that the variation of EoR 2D power spectrum caused by the frequency artifacts can be ignored. As a result, even a tiny ($\num{\sim 0.1}\%$) uncertainty of instrumental error or miscalibration can make the contamination of Galactic free--free emission become much stronger, particularly inside the EoR window. All the above analyses further support our conclusion that Galactic free--free emission is an important foreground component and needs to be carefully removed in the forthcoming EoR detections.

\subsection{Impacts on Component Separation}

The	Galactic free--free emission shows an impact on the separation of EoR signal, which can also be used to constrain the component separation of Galactic diffuse radiation (including both the synchrotron and free--free emissions; \citealt{Tegmark98,Delabrouille13}). The blind and the parametric fitting methods are the two major methods proposed to tackle the component separation problems (i.e., \citealt{Bennett03,de08,Delabrouille13,Planck-XII14,Zheng17,Thorne17}). The blind method (i.e., independent components analysis; ICA; \citealt{Maino07}) will be complicated by the Galactic free--free emission, especially in the power spectra $k$-space, since its power is more luminous than that of the EoR signal by about $3$ orders of magnitude. For the parametric fitting, the Planck Collaborations\footnote{\url{http://www/esa.int/Planck}} have provided a powerful algorithms, {\sc Commander}\footnote{\url{https://commander.bitbucket.io}}, which is for joint CMB estimation and component separation in parametric fitting via Gibbs sampling (\citealt{Planck-XII14,Planck-XXIII15,Planck-XXV15,Planck-X16,Planck-IX16,PlanckXIII16}). Moreover, the Galactic free--free emission acts to flatten the spectra index (i.e., from the original value \num{\sim -2.7} at \SI{150}{\MHz} to \num{\sim -2.5} due to the flat spectral index \num{\sim -2.1} of Galactic free--free emission) when adopting the parametric fitting method, so that an extra term (i.e., more parameters) should be needed to take this component into account.

\section{Summary}
\label{chap:summary}

We have evaluated the contamination of Galactic free--free emission on the EoR signal detection, for which we have incorporated the latest SKA1-Low layout configuration to take the instrumental effects into account. By comparing the power spectra between Galactic free--free emission and the EoR signal as well as Galactic synchrotron emission and masked extragalactic point sources in the \numrange{116}{124}, \numrange{146}{154}, and \SIrange{186}{194}{\MHz} frequency bands, we have shown that the Galactic free--free emission causes severe contamination on the EoR signal, especially toward lower frequencies (\SI{\sim 116}{\MHz}). In addition, we have estimated the effects of sky positions and frequency artifacts, both of which further support our conclusion that even inside the properly defined EoR window the Galactic free--free emission is still a non-negligible contaminating source and should be carefully dealt with in future EoR experiments. Moreover, we discuss the impact of Galactic free--free emission on the component separation, which will complicate the blind method and act to flatten the spectra index (\num{\sim -2.5}) for the parametric fitting method.

\section*{Acknowledgments}

The Virginia Tech Spectral-Line Survey (VTSS), the Southern H-Alpha Sky Survey Atlas (SHASSA), and the Wisconsin H-Alpha Mapper (WHAM) are all funded by the National Science Foundation (NSF). The SHASSA was operated by the Association of Universities for Research in Astronomy, Inc., under the cooperative agreement with the NSF, and the SHASSA observations were obtained at Cerro Tololo Inter-American Observatory. The WHAM facility was designed and built with the help of the University of Wisconsin Graduate School, Physical Sciences Lab, and Space Astronomy Lab, whose remote operation was provided by the NOAO staff at Kitt Peak and Cerro Tololo.
	
We gratefully acknowledge the reviewer for the constructive comments that greatly help improve the presentation of this work. We would like to thank M. G. Santos for providing the {\sc Simfast21} code\footnote{\url{https://github.com/mariogrs/Simfast21}}, Douglas P. Finkbeiner for providing the all-sky \Ha intensity map, and Fred Dulwich for providing the latest SKA1-Low layout configuration. All simulations are performed on the high-performance cluster at the Department of Astronomy, Shanghai Jiao Tong University. This work is supported by the Ministry of Science and Technology of China (grant nos. 2018YFA0404601)
and the National Natural Science Foundation of China (grant nos. 11621303, 11835009, 11973033).

\bsp	
\label{lastpage}
\end{document}